# Methods for the Reconstruction of Parallel Turbo Codes

Mathieu Cluzeau
INRIA-team SECRET

Matthieu Finiasz
ENSTA

Jean-Pierre Tillich
INRIA-team SECRET

*Abstract*—We present two new algorithms for the reconstruction of turbo codes from a noisy intercepted bitstream. With these algorithms, we were able to reconstruct various turbo codes with realistic parameter sizes. To the best of our knowledge, these are the first algorithms able to recover the whole permutation of a turbo code in the presence of high noise levels.

## Introduction

In this article, we focus on the problem of reconstructing a turbo code from a noisy intercepted bitstream. The code reconstruction problem consists for the eavesdropper in recovering the turbo code struture, or, more precisely, finding a decoding algorithm, ideally as efficient as the algorithm of the legitimate recipient. The problem of code reconstruction has already been addressed for a variety of codes [2]–[4], [7]–[9], [11], [13], [14]. Here, we focus on turbo codes, for which the problem has not yet been solved. The rest of this introduction is dedicated to recalling some basic definitions and notation on turbo codes. Sections I and II present two new methods for turbo code reconstruction and practical results we obtained. Then, Section III gives some insights on the benefit of combining these two methods and explains how to apply them to punctured turbo codes.

### A. Parallel Turbo Codes

We focus on reconstructing systematic parallel turbo codes of rate $\frac{1}{3}$ without puncturing. The input $X = \sum X_i D^i$ is split in blocks of length $N$ and is copied directly at the output. $X$ is also fed into a recursive convolutional encoder with encoding fraction $\frac{P'(D)}{Q'(D)}$. In other words, the output of this convolutional encoder is given by $Y = \sum Y_i D^i = \frac{P'(D)}{Q'(D)} X$. A certain permutation $\Pi$ (of length $N$) is applied to the input and it is then given to a second recursive convolutional encoder associated to the fraction $\frac{P(D)}{Q(D)}$. It yields the output $Z = \sum Z_i D^i = \frac{P(D)}{Q(D)} \sum X_{\pi(i)} D^i$, see Figure 1 for the whole scheme. Throughout this paper we consider that $M$ output blocks have been intercepted, corresponding to $MN$ input bits.

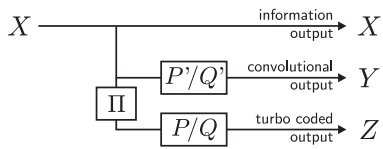

Figure 1. A systematic parallel turbo code with 3 outputs.

### B. The Problem of Turbo Code Reconstruction

Turbo code reconstruction requires to solve several problems, but many of them are already solved by (punctured) convolutional code reconstruction algorithms [7]. First, one needs to reconstruct the $P'/Q'$ convolutional encoder. Existing techniques allow to do this efficiently by looking for short dualwords of this convolutional code. Once this part is reconstructed, this reveals the interleaving and allows us to isolate the turbo coded output. At the same time, the knowledge of the convolutional encoder can help us decrease the noise level in the information output: this is not required for reconstruction, but can improve the performances of our algorithms.

Once the information and turbo coded outputs have been isolated, one needs to recover the polynomials $P$ and $Q$ used in the encoder and the permutation $\Pi$. The number of possibilities for $P$ and $Q$ are quite small as low degree polynomials are usually used for turbo codes, so this part can be solved by an exhaustive search. However, the size $N$ of the permutation is usually very large (up to 20 000), and most of the work consists in recovering $\Pi$. This is the problem we focus on.

*Previous Works:* The problem of reconstructing a permutation from a noisy output stream was first studied by Barbier in [2] where he gave an analysis of a simple algorithm. He focuses on the problem of recovering a permutation of size $N$ from a set of $N$-bit vectors and permuted versions of these vectors assuming some noise has been added. In the context of turbo code reconstruction, this algorithm cannot be directly applied: instead of a permuted version of the information, we have an *encoded* permuted version of the information. In order to apply Barbier's algorithm one first has to decode the turbo coded output. If there is no noise on the turbo coded output this is possible: one simply has to multiply it by $Q/P$ and then apply Barbier's algorithm. However, if there is noise on the turbo coded output this technique no longer applies.

Another approach by Côte and Sendrier [10] makes the opposite assumption: if there is no noise in the information output and some noise in the turbo coded output, they can recover the permutation $\Pi$. They generalize this approach to the case where there is noise in the information output but their technique requires a large amount of intercepted data, unless the noise is low or the permutation is short. This approach remains interesting as it is very efficient and, in practice, the noise level in the information output can be reduced using the convolutional output $Y$: it is thus natural to assume lower noise levels in the information than in the turbo coded output.

## I. Reconstruction Using Low Weight Dualwords

The algorithm we present here takes as input the two noisy output streams $X$ and $Z$ and outputs the polynomials $P$ and $Q$ and some parts (or the totality) of the permutation $\Pi$.

### A. Basic Idea

As for linear block codes, the concept of dualwords also exists for convolutional codes. A convolutional code is defined by a generator matrix of polynomials (or rational fractions) and its dual is simply the vector space spanned by polynomial vectors orthogonal to this matrix. Any element of this dual vector space is a dualword and to any such dualword corresponds a set of binary parity check equations.

For the turbo codes we consider, a systematic encoder of parameter $P/Q$, dualwords are of the form $(\lambda P, \lambda Q)$ where $\lambda$ is a polynomial. For example, for $P/Q = 1 + D^2 + D^3/1 + D + D^2$, the pair $(1 + D^2 + D^3, 1 + D + D^2)$ is a dualword (here $\lambda = 1$). Any parity check equation of the form $X_{\pi(i)} \oplus X_{\pi(i+1)} \oplus X_{\pi(i+3)} \oplus Z_{i+1} \oplus Z_{i+2} \oplus Z_{i+3}$ is thus equal to zero on the noiseless outputs $X$ and $Z$.

The main idea of this technique is to take advantage of such dualwords of very low weight (that is, weight 6 or 8) and exhaustively search for corresponding parity check equations. Each parity check equation we can find will give us several informations:

- the parity check bits on output $Z$ give us the polynomial $\lambda Q$ of the corresponding dualword,
- the number of parity check bits on $X$ give us the Hamming weight $w_0$ of $\lambda P$,
- once $P/Q$ has been guessed, the knowledge of $\lambda Q$ allows us to compute $\lambda P$ and the parity check bits on $X$ reveal a subset of $w_0$ positions of the permutation $\Pi$.

### B. Classification of the Polynomials

Each time we find a parity check for the intercepted output streams we can deduce a pair $(w_0, \lambda Q)$. Such a pair can only correspond to some encoders $P/Q$. We want to classify all encoders $P/Q$ (up to a given degree) depending on their dualwords and the associated $(w_0, \lambda Q)$ pairs. As this classification can be precomputed, we do not need a particularly efficient algorithm and simply use exhaustive search: we go through all polynomials $\lambda$ up to a certain degree, compute all multiples $(\lambda P, \lambda Q)$ and store all pairs $(w_0, \lambda Q)$ of low weight.

Allowing multiples of weight up to 6 is enough to classify uniquely all pairs $(P, Q)$ when $P$ and $Q$ are of degree 3 or less. Similarly, multiples of weight up to 8 are enough to classify uniquely all pairs of degree 5 or less, which will almost always be the case for a turbo code.

In order to use this classification[1], we now simply need to find parity check equations for the intercepted turbo code output. For example, suppose two dualwords have been found: $\lambda Q = 1 + D^2 + D^4$ with $w_0 = 3$ and $\lambda Q = 1 + D + D^5$ with $w_0 = 3$. The first dualword restricts the list of possible $P/Q$ to two candidates: $1 + D + D^3/1 + D + D^2$ and $1 + D^2 + D^3/1 + D + D^2$. The second dual word is enough to decide that $1 + D^2 + D^3/1 + D + D^2$ was indeed used.

### C. Finding Parity Check Equations

In order to look for parity check equations we build a matrix from the intercepted words. This matrix has $M$ lines (the $M$ intercepted blocks) and $2N$ columns, $N$ columns coming from the permuted $X$ and the $N$ others from $Z$. Searching for parity checks for this matrix is the exact same problem as for linear block code reconstruction. We can thus apply the techniques presented in [8]. Here, the block size $2N$ is usually very large and techniques based on the Canteaut-Chabaud algorithm [5] can only work for very low noise levels. We therefore prefer to use exhaustive search techniques like the Chose-Joux-Mitton algorithm [6] which focuses on very low weight parity checks and can thus tolerate much higher noise levels.

[1] We did not include the classification here because of its size but it can be easily recomputed.

The Chose-Joux-Mitton algorithm uses a generalized birthday algorithm technique to go through all parity check equations of a given weight $w$. It has a time complexity of $O(N^{\lceil \frac{w}{2} \rceil})$ and memory complexity $O(N^{\lceil \frac{w}{4} \rceil})$. We can thus use it to find parity check equations of weight 8 for values of $N$ up to a 1 000 or 2 000, but can only afford to look for parity checks of weight 6 for larger values (up to 20 000).

*Theoretical Analysis:* In order to measure the efficiency of this algorithm, we need to measure what proportion of the permutation it can recover, depending on the amount of noise. Note that in order to apply the Chose-Joux-Mitton algorithm, we need to make hard decisions on the values of the intercepted bits. We thus consider that the Gaussian channel of standard deviation $\sigma$ is transformed into a binary symmetric channel with cross-over probability $\tau$. In order to simplify the analysis we make two assumptions:

- what happens on the edges of a word is negligible: we suppose that any position of the permutation appears in $w_0$ shifts of a same dualword,
- any position of the permutation found in one dualword can be considered recovered: it can be found using an exhaustive search in $w_0!$ tries (which is negligible).

Thus, the quantity we need to measure in order to analyze our algorithm is the number of positions of the permutation which do not belong to any of the dualwords we found.

Using the same analysis as in [8], we can compute the probability $P_w$ of finding a given parity check of weight $w$ in a single run of the algorithm. We have: $P_w = (\frac{1+(1-2\tau)^w}{2})^\ell$ with $\ell = \frac{w}{2}(1 + \log_2 N)$.

Now, to determine the probability that a position of the permutation belongs to none of the parity checks the algorithm finds, we need to know the number of parity checks involving this position. This depends on the polynomials $P$ and $Q$: if the weight $w$ of the parity check is split in $w_0$ on the information part and $w_1$ on the turbo coded part, $w_0$ shifts of this parity check will involve a given position. So, the number of possible parity check for one position is the sum of the $w_0$ for all dualwords of $P/Q$ in our classification. We denote by $W$ this weight. Then, the number $N'$ of positions of the permutation that belong to no parity check equations after one run is $N' = N \times (1 - P_w)^W$. Of course, it is possible to decrease this $N'$ by running the Chose-Joux-Mitton algorithm several times on different windows of size $\ell$.

For example, for $P/Q = 1 + D^2 + D^3/1 + D + D^2$, with $w = 6$, there are 5 dualwords for a total weight on the information side of 15. For a length $N = 10\,000$ and $\sigma = 0.43$ resulting in $\tau = 0.01$, in one run of the algorithm $N' = 2\,868$ positions of the permutation should remain unfound and in two runs, only 823 positions should remain.

## II. RECONSTRUCTION USING STATISTICS ON THE CONVOLUTIONAL ENCODER ENTROPY

Similarly to the previous technique, this second algorithm takes the two noisy outputs $X$ and $Z$ as input and outputs the first positions of $\Pi$ (or all of it). However, here $P/Q$ needs to be known in order to use this algorithm. We therefore run the algorithm for all possible encoders $P/Q$ and only the correct choice will give an output. The algorithm will stop after a very small number of iterations for bad choices.

### A. Basic Idea

The key idea here is to make use of a distinguisher which is able to discriminate a typical decoding of a convolutional code over a given noisy channel from a decoding where all the bits come come from the noisy channel, except one which is completely random. Such a distinguisher can be used to recover recursively the permutation $\Pi$ in the following way. Assume that we have already found the $i-1$ first systematic positions of the convolutional encoder. We check now all remaining $N-i+1$ possible remaining positions for the $i$-th systematic bit by decoding the sequence formed by the $i$ first systematic positions and the associated redundancy. If we have at our disposal the aforementioned distinguisher we are able to find the value of the permutation because the right hypothesis for the $i$-th information bit corresponds to a typical decoding, whereas a wrong hypothesis corresponds to the second type of decoding. Such a distinguisher only exists if the $i$-th redundancy bit is not independent from the $i$-th systematic bit. This is the case when $P$ and $Q$ have a non-zero constant coefficient. This is always the case in practice and we make this assumption in this section.

We build a distinguisher from the BCJR decoding algorithm [1] which is precisely the decoding algorithm used to decode turbo codes. Consider the distribution during the BCJR algorithm of the forward probabilities $F_i$ on the state of the encoder when the $i$ first couples of noisy information bits and redundancy bits have been used. Consider also a related random variable $F'_i$ which corresponds to the forward probability of the BCJR algorithm if the decoder is fed with the $i-1$ first couples of noisy information and redundancy bits and then a couple formed by a random bit coming from the channel and the $i$-th noisy redundancy bit. Our aim is to find a distinguisher which after observing a sequence of samples which come all from realizations of $F_i$ or all from realizations of $F'_i$ tells us in which case we are. To simplify the task of finding a distinguisher for the BCJR decoding algorithm we are going to consider a one-dimensional statistic for $F_i$. We choose a functional which is invariant by permuting the states and which is basically a measure of how close $F_i$ is to the uniform distribution. The entropy meets these properties and it is our choice here. We denote by $H(F_i)$ the entropy of the distribution, that is $H(F_i) = \sum_a -F_i(a) \log F_i(a)$ where $F_i(a)$ stands for the forward probability that the $i$-th state of the encoder is $a$. We expect that the distribution of $H(F_i)$ will be sufficiently different from the distribution of $H(F'_i)$. This is the case as long the channel is not too noisy. Let us illustrate this with an example. Let us choose $P/Q = \frac{1+D^2}{1+D+D^2}$. We have plotted in Figure 2 the distribution of $H(F_i)$ and $H(F'_i)$ over a Gaussian channel for several values of the standard deviation $\sigma$ of the noise. When the noise is rather small ($\sigma = 0.8$) the two distributions are quite different. As could have been expected, the expected value of $H(F_i)$ is smaller than the expected value of $H(F'_i)$. The two distributions get closer to each other when the noise increases until being almost the same when the standard value of the noise exceeds 1.3. It should be noted that at this point the BCJR algorithm does not give useful information here for the extrinsic probabilities of a bit.

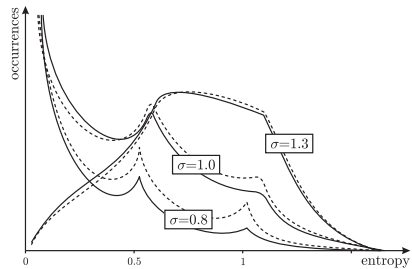

Figure 2. Distribution of the entropies $H(F_i)$ (plain line) and $H(F'_i)$ (dashed line) for Gaussian channels with standard deviations of 0.8, 1.0 or 1.3.

The fact that these two distributions are different when the noise is not too large leads us to the following method. First we perform for a given polynomial $P/Q$ a sampling test for obtaining an approximate distribution of quantized versions of $H(F_i)$ and $H(F'_i)$ for all values[2] of $i$ between 1 and $N$. Then we use a distinguisher for the two hypotheses (the samples stem from the distribution of $H(F_i)$ or from $H(F'_i)$). The one we will use here is just a Neyman-Pearson test using the approximate distributions of $H(F_i)$ and $H(F'_i)$.

### B. The Reconstruction Algorithm

Our algorithm works using a list of "candidates". Each candidate corresponds to a set of positions of the permutation (the positions that have already been guessed) and the $M$ internal state probability distributions associated with the $M$ received words. At the beginning, the list of candidates is initialized with one candidate (the systematic convolutional encoder $P/Q$ we are testing), it has 0 guessed positions and its internal states have probability 1 of being 0.

After $i-1$ steps of the algorithm, each candidate has $i-1$ guessed positions. After $N$ iterations, the permutation should be recovered (if the correct $P/Q$ was selected). Here is how the $i$-th step of the algorithm works for one candidate.

1) Assuming that $i-1$ positions of the permutation have been guessed, $N-i+1$ possibilities remain for the $i$-th position. Our candidate is thus split into $N-i+1$ possible candidates which we now have to filter.
2) for each possible candidate, we update the $M$ internal state probability distributions using the bit from the guessed $i$-th position. We then compute the entropy of each of these $M$ internal states. When the number of states is $2^m$, these entropies have values between 0 and $m$ so we split this interval in $w$ sub-intervals to obtain an estimate of the entropy distribution $D^{cand}$ of our candidate. We denote by $(D_j^{cand})_{j \in [0, w-1]}$ the fraction of words yielding an entropy in $[\frac{jm}{w}; \frac{(j+1)m}{w}]$.

---

[2]The complexity of this step can be reduced a little bit if we assume that the distribution of $H(F_i)$ and $H(F'_i)$ becomes stationary after a while.

3) this distribution $D^{cand}$ has to be compared with two target distributions $D^{good}$ and $D^{bad}$ which were precomputed using the same technique as in step 2 on a large set of generated samples: $D^{good}$ corresponds to samples of $H(F_i)$, $D^{bad}$ to samples of $H(F_i')$. We use a Neyman-Pearson test to decide which of $D^{good}$ or $D^{bad}$ is closer to $D^{cand}$. We thus compute the three values:

$T_{good} = \sum_{j=0}^{w-1} D_j^{good} \times (\log_2(D_j^{good}) - \log_2(D_j^{bad}))$,
$T_{bad} = \sum_{j=0}^{w-1} D_j^{bad} \times (\log_2(D_j^{good}) - \log_2(D_j^{bad}))$,
$T_{cand} = \sum_{j=0}^{w-1} D_j^{cand} \times (\log_2(D_j^{good}) - \log_2(D_j^{bad}))$.

4) we set a threshold $T$ close to $\frac{1}{2}(T_{good} + T_{bad})$ (see section II-C1 for more details on how to compute $T$). If $T_{cand} > T$ then the distribution $D^{cand}$ is closer to $D^{good}$ and we keep this candidate. Otherwise we discard it.

### C. Theoretical Analysis

*1) Threshold selection:* The selection of the threshold $T$ depends on several factors: the two distributions $D^{good}$ and $D^{bad}$ and two probabilities $\alpha$ and $\beta$ of respectively keeping a bad candidate and discarding the good one. Of course, depending on the number $M$ of available samples, some values of $\alpha$ and $\beta$ might not be possible to achieve.

To compute precisely the threshold $T$, we need to know the distribution of $T_{cand}$, computed either for good candidates or bad candidates. We denote by $z^{good}$, respectively $z^{bad}$, the random value taken by $T_{cand}$ when the candidates are drawn according to $D_{good}$, respectively according to $D_{bad}$. $\beta$ is defined by $\beta \stackrel{\text{def}}{=} P[z^{good} < T]$, whereas $\alpha$ is given by $\alpha \stackrel{\text{def}}{=} P[z^{bad} > T]$.

In order to compute estimates of $\alpha$ and $\beta$ we use the large deviations estimates of [12, Appendix 5A]. For this purpose, we introduce the random variable $X^{good} \stackrel{\text{def}}{=} \log_2 \frac{D^{good}(H)}{D^{bad}(H)}$. Here $H$ stands for a random variable drawn according to the distribution $D^{good}$ and $D^{good}(H) = D_j^{good}$ where $j$ is such that $H$ belongs to the $j$-th quantization interval, that is $[\frac{jm}{w}; \frac{(j+1)m}{w}]$. We define in the same way $D^{bad}(H)$. We also define $X^{bad}$ similarly with $H$ being now a random variable distributed according to $D^{bad}$. We let $\mu_{good}(s) \stackrel{\text{def}}{=} \ln \mathbb{E}(e^{sX^{good}})$ and $\mu_{bad}(s) \stackrel{\text{def}}{=} \ln \mathbb{E}(e^{sX^{bad}})$. Denoting by $\mu'$ and $\mu''$ the first and second derivatives in $s$ of these functions, we have the following formulas:

$$P[z^{bad} > \mu'_{bad}(s)] \simeq \frac{e^{(\mu_{bad}(s) - s\mu'_{bad}(s))M}}{|s|\sqrt{2\pi M \mu''_{bad}(s)}} (= A(s, M)),$$

$$P[z^{good} < \mu'_{good}(s)] \simeq \frac{e^{(\mu_{good}(s) - s\mu'_{good}(s))M}}{|s|\sqrt{2\pi M \mu''_{good}(s)}} (= B(s, M)).$$

We use these equations in order to find the suitable threshold $T$ and choose the smallest $M$ such that there exists $s_\alpha$ and $s_\beta$ satisfying:

$$\begin{cases} A(s_\alpha, M) \leq \alpha \text{ and } B(s_\beta, M) \leq \beta \\ \mu'_{good}(s_\beta) = \mu'_{bad}(s_\alpha) \quad (= T) \end{cases}$$

*2) Complexity Analysis:* The complexity of each step of the algorithm is linear in the number of candidates. As long as the number of candidates remains close to one, the complexity of the algorithm is polynomial. However, a bad choice of the probability $\alpha$ can lead to an exponential number of candidates. For one candidate in the $i$-th step of the algorithm, one has to test $N - i + 1$ possible positions and for each of them compute the value $T_{cand}$. Generating $D^{cand}$ costs $O(M2^m)$ and computing $T_{cand}$ from it costs $O(w)$. The sampling fineness $w$ is a constant so we can ignore it. The cost of one step of the algorithm is thus $O(NM2^m)$ for each candidate and the overall complexity of the algorithm is therefore $O(N^2M2^m)$ if the number of candidates remains bounded.

### D. Experimental Results

We have run several tests using a turbo code of parameters $P/Q = \frac{1+D^2}{1+D+D^2}$ and various permutation sizes and noise levels. For all our tests, we chose $\alpha = 1/N$ so as to keep the number of bad candidates close to one, and $\beta = 0.01/N$ for a probability of discarding the correct candidate before the end of the $N$ steps of the algorithm close to 1%. Here are some of the results we observed.

- If we choose $M$ and $T$ as given by our theoretical analysis, the algorithm runs smoothly and outputs a single candidate: the correct one.
- When running the algorithm for the wrong encoder $P/Q$, the algorithm starts with a single candidate and discards it after only a few steps.
- The distributions $D^{good}$ and $D^{bad}$ are computed assuming that an error occurred (or not) on the last position. Once a bad candidate has been selected these distributions are not optimal to eliminate it at the following step: a distribution ending with two errors would be better. In practice, this did not cause any problem and bad candidates never generate further bad candidates.
- All our test were made using Gaussian noise. We did not implement any other noise model but the formula we use to compute $M$ and $T$ are true for any distribution. Other noise models might thus require to use a much larger number of intercepted words but should work as well.

| $N$ | $\sigma$ | $M$ | (theory) | running time in seconds | proportion of the permutation recovered |
|---|---|---|---|---|---|
| 64 | 0.43 | 50 | (48) | 0.2 | 100% |
| 64 | 0.6 | 115 | (115) | 0.3 | 100% |
| 64 | 1 | 1380 | (1380) | 12 | 100% |
| 512 | 0.6 | 170 | (169) | 11 | 100% |
| 512 | 0.8 | 600 | (597) | 37 | 100% |
| 512 | 1 | 2800 | (2736) | 173 | 100% |
| 512 | 1.1 | 3840 | (3837) | 357 | 100% |
| 512 | 1.3 | 29500 | (29448) | 4477 | 100% |
| 10000 | 0.43 | 300 | (163) | 8173 | 100% |
| 10000 | 0.6 | 250 | (249) | 7043 | 100% |

As we can see in the previous table, we manage to reconstruct the whole permutation even for extreme noise level of $\sigma = 1.3$. For small parameters the algorithm is very fast as the number of candidates at each step remains close to 1. However, the time difference between noise levels of 1.1 and 1 for a permutation of length 512 is larger than expected. This comes from the fact that at several steps the number of candidates increases to $\sim 20$. We also tried to run our algorithm with less words than theoretically required. For a permutation of length 64 and a noise level of 0.6, only 60 words (compared to the 115 theoretical bound) were enough to recover the whole permutation in 70% of our tests.

Concerning longer permutations, as the complexity is quadratic in $N$ and linear in $M$, our algorithm still performs as expected but takes much longer. Recovering a permutation of size 10 000 takes a few hours. This shows that our algorithm is able to reconstruct any non-punctured parallel turbo code.

III. FURTHER IMPROVEMENTS

A. Combination of the Two Methods

One can combine two algorithms we presented to improve their performances. For instance, the first algorithm using low weight dualwords can be used to guess $P/Q$ without the need for an exhaustive search. However, recovering the whole permutation with it can require a large number of runs depending on the noise level. In particular, the last positions of the permutation are more expensive to guess than the first ones. The second algorithm using entropy statistics behaves the opposite way: the first positions of the permutation are more expensive to recover as more choices are possible. A good procedure is the following:

1) Run the first algorithm once. If no dualwords were found or only very few were found, the noise level is probably too high for this algorithm. Jump directly to step 3.
2) If several dualwords were found in step 1, then there are probably enough dualwords to deduce $P/Q$ from the classification. Also, each further run of the first algorithm can help reduce the number of unknown positions in the permutation. Run this algorithm again a few times, as long as each additional run significantly decreases the remaining number $N'$ of positions of $\Pi$.
3) Except if the noise level is very low and the first algorithm was able to reconstruct the whole permutation, the reconstruction should always end with a run of the second algorithm. This algorithm is relatively slow if $N'$ is large, but will almost always recover the whole permutation if enough intercepted words are available.

B. Puncturing

The algorithms we presented were for parallel turbo codes with no puncturing: we assumed that all the bits of outputs $X$ and $Z$ were intercepted. Most real world turbo codes however use puncturing in order to obtain different transmission rates: typically, only one out of two bits of the outputs $Y$ and $Z$ are transmitted. In such a case, both our methods can be adapted.

For our first method, when a puncturing is applied, the dualwords involving punctured bits no longer exist. Our algorithm will still work, but only output dualwords with no punctured bits. This has two consequences: first, the number of available dualwords is smaller and the probability of not finding any dualwords for one position is thus higher, secondly, the classification is be sparser and identifying $P/Q$ requires to look for higher weight dualwords, increasing the cost of the exhaustive search.

For our second method, puncturing also increases the cost of the algorithm. Depending on the puncturing pattern and the polynomials $P/Q$, each intercepted bit of $Z$ depends on 1 or more positions of $\Pi$. For a puncturing of rate $\frac{1}{2}$, each output bit depends on the 2 previous bits of the permuted $X$. Therefore, instead of guessing each position of $\Pi$ independently, we have to guess 2 positions (or more for sparser puncturing) at a time. The complexity of the algorithm will thus increase from $O(N^2)$ to $O(N^3)$. Also, instead of having to guess between two possible entropy distributions, four distributions are possible depending on the exactness of both guessed bits. This makes the analysis of this algorithm more complex: the problem of distinguishing between two distributions is well known and classical results allow us to make optimal decisions. For four distributions this is no longer the case, especially if distributions we consider are not independent.